# Over-reliant and underprepared: First responder resilience to disruption of operational technology

**James Connor** **Courtney Crooks** **Ryan Shandler**

*Georgia Institute of Technology*
*jconnor@gatech.edu*

## Abstract

The Emergency Services Sector (ESS) is responsible for taking life-saving action in the face of extreme crises. To enhance its response capabilities, the sector has spent the last two decades investing in new technology to improve their operational efficiency. This has created significant interdependencies and a culture among first responders of technical overreliance, raising questions about what would occur if those tools were suddenly disrupted. Several works have discussed the potential for terrorism to intentionally target ESS technology as part of a larger kinetic attack. Despite this, the resilience of first responder agencies to technological disruption and their preparedness for such an attack has largely remained unexplored. To bridge this gap, we implemented a novel scenario-based interview methodology involving senior officials from emergency service agencies across the United States. These interviews explored how each agency prepared for a disruption of operational technology and what factors might influence different levels of preparation between them. Participants were asked to describe their responses to a hypothetical scenario modeled after historical cyber-attacks against ESS infrastructure. The results indicate that while ESS agencies are prepared for short-term technological disruptions, they are likely to struggle with long-term or hybrid incidents. The data further revealed that agencies often lack dedicated expertise in cybersecurity, technology, and emergency management that could otherwise elevate resilience to interruptions of operational technology. These findings emphasize the need for investment in redundant analog systems, increased training for frontline personnel, and a focused effort against loss of specialized technical expertise among ESS agencies.

## 1. Introduction

In 2023, the Louisville MetroSafe 911 Communications Center received reports that a dirty bomb had detonated at a local entertainment venue, followed by internal communications that they had been breached by a cyber-attack [1]. Facing degraded technological conditions, officials expressed uncertainty over their ability to continue operations. This incident was merely a simulated exercise known as "Operation Red Dawn," yet it underscored a persistent and well-documented concern in both policy circles and academic literature: *first responder operations are vulnerable to technical disruption*.

ESS is a division of critical infrastructure (CI) focused on protecting life and property. The five branches of ESS consists of fire and rescue services, law enforcement, emergency medical services (EMS), emergency management, and public works, with each branch adopting specialized functions and objectives in support of overall public safety [2]. To support this mission, ESS agencies have increasingly integrated advanced technologies into their operations, including body cameras, cardiac monitors, digital emergency operation centers (EOCs), and myriad networked devices that improve situational awareness. These technologies have enhanced efficiency and response capability, yet they have simultaneously introduced systemic risk. Expanded technological adoption increases exposure to security vulnerabilities, while overdependence on such tools can amplify the harms should they be exploited. Given the time-sensitive nature of emergency response, the risks associated with technical disruption are particularly salient.

Although ESS personnel are thoroughly trained to deal with uncertain situations, their increasing dependence on vulnerable digital systems poses questions about the resilience of the sector to mounting digital threats. While the nature of such technical disruptions has been theorized, there is little evidence documenting how ESS agencies would respond to a genuine instance of Operation Red Dawn. This paper seeks to address this gap by empirically investigating several core questions:

- RQ1. How prepared is ESS for a cyber-based disruption?
- RQ2. How would emergency service agencies respond to such disruption during a large-scale incident, specifically in a terrorism context?
- RQ3. What factors are most central in guiding successful responses among first responders?

To answer these questions, the paper proceeds by first evaluating the extant theories in critical infrastructure and cybersecurity literature as they pertain to the resilience of emergency services facing technical disruption. With this data, we conduct a series of dynamic scenario-based interviews centered on high-level simulations of such disruption with emergency personnel across six states. The resulting data revealed effectiveness for continuity-planning among participants, but simultaneously offered confirmation of overdependence on technology and a lack of preparedness among interviewed personnel should their digital systems fail. We conclude by discussing the policy implications and practical steps required to enhance ESS resilience, as well as future opportunities for generalizing these exploratory findings and furthering research for this underexplored field.

## 2. Literature Review

### *2.1. The Emergency Services Sector & Technological Dependencies*

ESS agencies are known for their flexibility and being able to handle any incident that might occur, regardless of complexity or novelty. This phenomenon is known as comprehensive emergency management (CEM), although first responders often label it as the “all-hazards” approach to emergency preparedness [3,4]. CEM is based in the pursuit of resilience within an emergency system, uniting the common response characteristics for different types of hazards while acknowledging the need for planning based on locally identified risks [2,5,6]. Thus, the common practice of developing pre-written standard operating procedures to manage a complex incident is balanced with the novel context that an incident exists within and the need to adapt to the unprecedented.

Regardless, there still exists an inherent degree of flexibility within ESS that is accompanied by a significant heterogeneity among agencies, particularly those belonging to the same ESS branch. For instance, fire departments in areas with a large population are more likely to feature a paid staffing model, but tend to have less apparatus or personnel relative to their population size [7]. Their capabilities differ as well: urban EMS agencies with shorter response times tend to be slower in adopting blood transfusion capabilities than their rural counterparts [8]. This phenomenon further extends to cybersecurity protections, in which small municipalities and rural critical infrastructure systems are less likely to significantly invest in cybersecurity compared to their urban, metropolitan counterparts, largely as a result of strategic threat modeling and a lack of resources [9,10]. Whether planning for hazards, developing specialized resources, or contributing to cybersecurity protections, ESS agencies differ significantly as a result of the variety that exists among jurisdictions and the hazards they prioritize.

Despite their ability to respond to all emergencies, ESS is not a self-sufficient sector and it maintains complex and interdependent relationships with other areas of critical infrastructure [11]. A 2014 DHS report emphasized the interdependence of ESS with various sectors including energy, water, and healthcare, but further identified the communications sector as the sole dependency [12] (although some authors contend that communications is also categorizable as an interdependency) [13,14]. Regardless of the classification, it remains clear that ESS is dependent on the communications sector because it enables the technology that emergency personnel increasingly rely on. Public safety answering points (PSAPs) rely on the sector to ensure widespread accessibility and accurate routing of 911 calls. Frontline responders work with radio equipment to facilitate intra- and inter-agency coordination. Mobile data terminals (MDTs), computer aided dispatch (CAD), automatic vehicle locators (AVLs), and electronic patient care reports (ePCRs) are all used to transfer pertinent information between dispatchers and units over existing telecommunication networks. Even simple administrative tasks, such as staffing and inventory management, are increasingly outsourced to internet-based services—services that are only usable if the underlying infrastructure is available.

The modern responder employs a vast array of technology, networked or otherwise. However, the effectiveness it introduces has been accompanied by a new dependence among ESS personnel, not on a particular sector, but on technology itself [2,15]. Dawkins et. al conducted a cross-branch assessment of first responders and their opinions on communication technology—one of the recurring concerns raised across both interviews and open-ended survey questions was that ESS had become "too dependent" on technology [16]. As one respondent from the study explained, overreliance can cause "[first responders] to freeze when technology fails. And it does eventually fail." As ESS continues to innovate—introducing drones, body cameras, telehealth, wireless smoke detectors, NG911, and other systems—the underlying question shifts from *if* the technology will fail to how first responders will react *when* the technology fails.

### *2.2. Hacking the Emergency Services*

New technology exposes ESS by introducing a wider attack surface to be targeted by hostile actors, an exposure that is exacerbated by the sector's dependence on that same disruptable technology [2]. Along with each new device or software introduced to first responders comes the prospect of it being targeted, putting operations at risk [17]. There are multiple categories of attack vectors that have been materially exploited against first responder technologies. Table 1 presents a taxonomy of cyber-attacks against ESS that have been documented across existing literature and available news reporting. Appendix A supplements this taxonomy by presenting historical instances of each attack vector included in the table. With the expanding adoption of NG911, first responder drones, and other technology, it is likely that the list will continue to grow [18].

**Table 1: Taxonomy of historical cyber-attacks targeting ESS**

| Category | Description |
|---|---|
| Telephony-based Denial-of-Service (TDoS) | Numerous illegitimate connections to 911 are facilitated by an attacker to disrupt access for legitimate callers [19]. |
| Ransomware | The execution of malware on a system for the purpose of encrypting data and potentially coercing a monetary ransom. This is particularly disruptive when targeted toward PSAPs and CAD systems [20,21]. |
| Data Breaches / Doxxing | The unauthorized access and potential disclosure of sensitive or personally identifiable information, especially about ESS personnel [22,23]. |
| Insider System Compromise | The exploitation of legitimate access by ESS personnel to disrupt a system or provide access to other malicious actors [24]. |
| Radio Compromise | It has been demonstrated by researchers that radio communications can be eavesdropped or forged through exploitation of vulnerabilities in certain radio standards [25]. Historical incidents have also demonstrated the ability to interfere with first responder radio traffic via jamming [26]. |
| Emergency Alert System (EAS) Intrusion | The compromise of an EAS device that results in the distribution of a false emergency alert. This has been seen in multiple forms, including siren activations, television alerts, and text message notifications [27]. Such acts can induce significant panic among the general populace and result in their contacting emergency services to obtain credible information. |
| Swatting | The dislocation of first responders through the false reporting of an emergency. Reports are typically designed to elicit a significant emergency response, including SWAT teams or bomb squads, and are often coupled with spoofing and doxxing to anonymize the perpetrator or target a particular individual [28]. |

Some of these vectors require greater technical expertise—such as radio compromise, TDoS, or EAS intrusion—while others simply need legitimate access or a telephone. The intended effects too can range, including disruptions of day-to-day responses, the generation of a new incident, or operational security impacts for frontline personnel. There exist myriad approaches for classifying such cyber-attacks against ESS infrastructure—prior work by Loukas and colleagues seemingly reached the same conclusion when they developed three distinct perspectives in their taxonomy that specifically covered cyber-physical EM systems [29]. Thus, this table does not serve as an exhaustive taxonomy, but as a systematic approach toward understanding the spectrum of digital threats that currently exist toward first responders.

The existence of these digital vulnerabilities does not mean that first responders should revert to entirely analog methods of communication and operation; these technologies have meaningfully increased ESS efficiency and contributed to a better equipped service [2]. However, the reality remains that cyber-attacks have the potential to negatively impact emergency services across multiple fronts, including response times and situational awareness. A handicapped PSAP intuitively takes longer to dispatch appropriate units due to the use of pen-and-paper, while a crashed telephone network creates extended time between an emergency occurring and 911 callers connecting with a dispatcher [30]. Responders displaced by swatting can suffer from a longer response distance. Impediments on radio communication can make it difficult for various ESS branches and personnel to coordinate their responses, allocate resources as needed, and ensure all first responders receive necessary information. In a field where time makes the difference for saving life and property, these harms become immediately tangible.

Not all cyber-attacks are accompanied by the same amount of risk. There are some attacks that might occur regularly, but are low-risk because they have a negligible impact on ESS operations. Conversely, there are certain categories of attacks that are incredibly damaging to operations yet remain low-risk because of the rarity with which they occur. Using this approach, the attack vectors described above can be organized in a risk matrix as displayed in Figure 1. Likelihood in this figure ranges from "Low," meaning the attack has not historically occurred outside a controlled research environment, to "High," meaning the attack has been regularly observed and requires little to no technical expertise to perpetrate. Severity in turn ranges from "Low," indicating little to no disruption of ESS operations, to "High," indicating widespread disruption with a complete breakdown in operational capability. For instance, a large TDoS attack was rated with a medium likelihood as it has existed outside of a theoretical context, but requires access to a large number of compromised devices that renders it challenging to execute [19]. Meanwhile, the same attack was rated with medium impact, as it would undermine ESS operations by disrupting 911, but would not prevent ESS personnel from effectively responding to emergencies. Our analytical approach to assigning risk and probability is described in full in Appendix A.

**Likelihood of Occurrence**

| Impact Severity | Low | Mod-Low | Moderate | Mod-High | High |
|---|---|---|---|---|---|
| **Low** | | | Website Defacement | | |
| **Mod-Low** | | | Data Breach<br>EAS Intrusion | | Swatting |
| **Moderate** | Radio Eavesdropping | Ransomware on CAD | Ransomware on PSAP | Data Breach (of LEO PII) | |
| **Mod-High** | Radio Jamming | | Large TDoS Attack | | |
| **High** | Radio Forgery | | | | |

**Fig. 1.** A risk matrix depicting different cyber-attacks targeting ESS and how they differ in likelihood and severity.

### *2.3. Terrorism and Cyber-Attacks*

The question naturally follows: who would be motivated to target ESS in cyberspace? Rowland and colleagues provide a framework to begin considering that question [31]. The illegality of attacking ESS rules out legitimate cyber actors, including corporations, non-profits, and all but the most extreme of activist groups. Profit motivations could encourage targeting by cybercriminals while ideological goals might drive interest among terrorists and hacktivists. Historical instances of cyber-attacks against ESS have been dominated by cybercriminals, pranksters, and nation-state actors [32–34], yet an increasing body of literature places particular concern toward extremist groups [35,36].

Terrorism is predominantly focused on targeting the general populace, a by-product of its interest in instilling widespread fear for the sake of political or societal coercion [37,38]. Targeting critical infrastructure does not induce the same casualties one can expect from bombing a densely populated event or tourist attraction. Instead, attacks against CI are focused on the strategic generation of terror via higher-order effects, using the disruption of essential services to create short-term fear and long-term institutional distrust [39,40]. While terrorism targeted against first responders maintains concerns for the casualties it might cause, a far greater danger consequently lies in the crippling of emergency services when they are needed most. The September 11th attacks and Hurricane Katrina featured suboptimal responses by emergency services due to communication failures, lack of situational awareness, and overwhelmed resources [30,41]. The results were telling: New Orleans residents reported post-disaster distrust in public health and law enforcement authorities [42], while the shortcomings of September 11th forced a national redesign of ESS and how it responds in cases of terrorist attacks [43].

Since 1970, there have been 226 recorded instances of terror attacks striking EMS and fire-rescue personnel [44,45]. Approximately 28% of those have been secondary attacks such as bombings and assaults directed at emergency services as they respond to an initial attack. However, the labeling of these attacks as secondary imprecisely invokes an image of chronological succession; it suggests that first responders will be targeted only after an explosion or shooting occurs. An assessment by DHS highlighted the implicit error: "cyberdisruption . . . during an emergency operation could dramatically disrupt or delay the initial response to an event" [12]. The report raises the cogent possibility of first responders being targeted before an attack begins, amplifying the kinetic results in an antecedent manner.

The possibility of terror actors targeting ESS technology has been suggested across multiple works [22,28,46–48], with one article by Lewis emphasizing the national security risk posed by terrorists using cyber capabilities "as corollaries to more traditional modes of attack" [49]. The allure of this strategy for terrorism is not insignificant given such a scenario's potential as a destabilizer, fear inducer, and force multiplier. In other words, there exists a rational motivation for extremist groups to direct cyber campaigns against ESS and its operational capability.

However, this concern exists in the shadow of continued debate over whether cyberterrorism is feasible. Researchers have cited a high barrier to entry, a lack of technical expertise among terror actors, and few historical instances as reasons to conclude that the threat of cyberterrorism might be exaggerated [24,50]. It is argued that terrorists would not resort to a new methodology when proven analog tactics continue to exact fear among the public [51], and that if a "cyber Pearl Harbor" [52] has truly been around the corner for the past decade, more attacks would have already occurred [53]. Yet evidence exists of traditional extremist groups developing in-house cyber capabilities [54–56], even as other theories have warned that the increased convenience of cybercrime-as-a-service may reduce the skillset required for terror actors to go digital [57]. Furthermore, there is an increasing sense that, depending on the definition subscribed to, acts of cyberterror may have already occurred [58–61]. Criticisms of the potential for cyberterrorism may be well-founded, but ignoring the possibility of cyber-attacks perpetrated by extremist groups could represent the same category of imagination failure that contributed to the unanticipated nature of the September 11th attacks [30].

## 3. Methodology

It is difficult to objectively measure the resilience of ESS agencies in the face of cyber-attacks that cripple their critical operations technologies. To rigorously measure this ability, we set in motion a novel, empirical methodology that recorded data from eight leadership-level personnel representing ESS agencies across the United States. We employed a scenario-based interview method, positioned between traditional vignette studies and full-scale table-top exercises, in which participants were presented with a detailed, realistic, fictional event and asked to describe how they would respond to the evolving incident. This scenario-based approach was further supplemented by a semi-structured interview that offered insight into the preparedness of each agency, and the expected response from other ESS personnel within their emergency response system.

The research team began by identifying individuals employed by emergency service agencies in areas likely to be targeted by terror actors. Identified individuals were then interviewed regarding their agency's preparation for a cyber-supported kinetic terror attack. Finally, the results were coded and interpreted using a thematic analytical framework to determine broader sector-wide conclusions for the emergency services.

### *3.1. Recruitment of Interviewees*

Terrorists do not uniformly target jurisdictions across the United States. Their interest in inspiring terror encourages target selection that maximizes visibility and casualties [62]. Therefore, in the interest of eliciting meaningful results, sampling was focused on emergency service agencies in local or regional municipalities defined by 1) a large population, 2) a history of being targeted by terror actors, or 3) a particular sensitivity to terrorism, in alignment with prior work by Riley & Hoffman [63].

To narrow the population of all emergency services across the U.S. based on population, the fifty states were grouped based on their U.S. census region: Northeast, Midwest, West, South, and Pacific. The two cities with the largest census population from each region were selected, with a restriction added that both cities could not be from the same state in order to encourage sample diversity [64]. The cities selected by population are displayed in Table 2.

**Table 2. Cities sampled based on population**

| Northeast | Midwest | West | South | Pacific |
|---|---|---|---|---|
| New York, NY | Chicago, IL | Los Angeles, CA | Houston, TX | Honolulu, HI |
| Philadelphia, PA | Columbus, OH | Phoenix, AZ | Jacksonville, FL | Anchorage, AK |

To supplement the population-based sample, additional cities were selected based on their having a history of terrorism or being locations commonly viewed as terror targets. Cities matching these criteria were derived from entries in the Global Terrorism Database [65]. Prior work by Piegorsch, Cutter, & Hardisty revealed 19 cities considered significantly vulnerable to a casualty-inducing terror attack [66]. Snowball sampling was also used based on recommendation from officials contacted during interview recruitment. This recruitment strategy resulted in the selection of 18 cities as displayed in Table 3.

**Table 3. Cities sampled based on sensitivity to or history of terrorism**

| **Sensitivity to Terrorism** * | **History of Terrorism** |
|---|---|
| 1. Atlanta, GA | 1. Boston, MA |
| 2. Baltimore, MD | 2. Las Vegas, NV |
| 3. Baton Rouge, LA | 3. Washington, DC |
| 4. Boise, ID | |
| 5. Charleston, SC | |
| 6. Charlotte, NC | |
| 7. Cleveland, OH | |
| 8. Columbia, SC | |
| 9. Louisville, KY ** | |
| 10. New Orleans, LA *** | |
| 11. Norfolk, VA | |
| 12. Raleigh, NC | |
| 13. Richmond, VA | |
| 14. St. Louis, MO | |
| 15. Tampa, FL | |

* Selected from Piegorsch et al. unless otherwise noted
** Snowball sampling
*** New Orleans was not included in the history column, as the 2025 Bourbon Street terror attack occurred after sampling.

For all cities, one agency from each ESS branch was selected. Public works was excluded both due to the branch not being typically involved in immediate response to terrorism and as a means of minimizing sample oversaturation for a particular city. For each sampled agency, researchers attempted to identify an interviewee holding a position in information technology, cybersecurity, emergency preparedness, or homeland security. Secondary preference was placed on personnel in an operational or executive role.

### *3.2. Scenario Construction and Interview Execution*

On the basis of the literature, we constructed a realistic scenario to reflect the attack matrix depicted in Figure 1. In alignment with the goal of developing a worst-case scenario, we selected three attacks to include: the attack with the most severe impact from each of the low, medium, and high likelihood ratings. Notably, radio forgery was excluded from consideration, as we were unable to identify its use against ESS outside of controlled conditions. These attacks were then combined with the context of a traditional kinetic terror attack to produce the final scenario that was presented to participants.

The final scenario consisted of an unspecified type of kinetic attack perpetrated by an unspecified threat actor (UTA). Prior to that attack, the UTA was said to have engaged in swatting to displace first responders from the planned locations of their kinetic attack. The UTA was then said to have launched a TDoS attack targeting 911 infrastructure to delay legitimate reporting of the kinetic attack and hamper situational awareness efforts. Finally, the UTA was said to have engaged in limited radio jamming at one of the kinetic attack sites to further undermine situational awareness and disrupt attempts at establishing effective incident command. While modeled after actual incidents and security advisories, the true testament to the scenario's legitimacy came in the form of several interviews during which participants [N1, N3, N4, N6] noted its realistic nature and their worries over it occurring. The full scenario is detailed in Appendix B.

The interviews were semi-structured and conducted via video-conferencing software. The first half of each interview was spent gathering responses to the questionnaire detailed in Appendix C, while the second half used the scenario to ascertain each interviewee's response to technological disruption. Interviews lasted for approximately one hour each, with one outlier [N1] of 24 minutes. Interviews were recorded and transcribed, with transcriptions then being manually reviewed and edited for accuracy. Data collection was terminated after the eighth interview, as researchers believed that there were no additional participants who would 1) meet the inclusion criteria, 2) be responsive to recruitment efforts, and 3) offer novel responses.

### *3.3. Data Analysis*

A Braun and Clarke thematic approach was used to analyze the data [67,68]. The lead author first reviewed each transcript to support data familiarization, while simultaneously recording initial observations and notable patterns. An inductive coding approach was then adopted, developing codes based on extracted quotations. Each quotation was distilled into keywords, which were used to synthesize a set of twenty-eight initial codes based on their expression of similar sentiments. Data immersion, frequency analysis, and reevaluation in the context of the underlying research questions were used to refine codes over several iterations, resulting in a final set of twenty-five codes that were applicable across the transcripts. Codes were reviewed by the second author to confirm findings, with disagreements resolved via discussion. A full list of themes, their codes, the number of transcripts those codes were present in, and their raw occurrence count across extracted quotations is listed in Table 4.

**Table 4. Final themes and their respective codes**

| Theme / Subthemes | | Codes | Transcript Frequency (%) | Quotation Count |
|---|---|---|---|---|
| Continuity in the Face of Technical Disruptions | | Confidence | 7 (87.5) | 14 |
| | | Coordination | 8 (100.0) | 28 |
| | | Public harms/good | 6 (75.0) | 11 |
| | | Contingencies | 8 (100.0) | 58 |
| | | Historical technical disruptions | 7 (87.5) | 19 |
| | | Importance of communications | 4 (50.0) | 9 |
| | | Mission-oriented continuity | 3 (37.5) | 4 |
| Planning for Failure | | Kinetic incident preparedness | 7 (87.5) | 29 |
| | | Worst-case planning | 4 (50.0) | 7 |
| | | Worst-case planning avoidance | 3 (37.5) | 3 |
| | | Agency harms | 7 (87.5) | 11 |
| | | Training/Exercises | 6 (75.0) | 10 |
| Frontline vs. Command-oriented Approach to Technical Disruption | | Procedural approach | 7 (87.5) | 21 |
| | | Flexible approach | 8 (100.0) | 26 |
| | | Procedurally flexible approach | 1 (12.5) | 2 |
| | | Technological overreliance | 5 (62.5) | 9 |
| | | Institutional knowledge | 3 (37.5) | 7 |
| Agency Partnerships and Expertise | Outsourcing of expertise | Outsourced expertise | 6 (75.0) | 32 |
| | | In-house expertise | 4 (50.0) | 7 |
| | | Lack of expertise | 5 (62.5) | 13 |
| | Relationships with Partners | Mutual-aid | 5 (62.5) | 11 |
| | | Interoperability | 5 (62.5) | 11 |
| | | Good relationship | 4 (50.0) | 6 |

Several initial themes were developed based on grouping similar codes together. Those initial themes were validated by comparing them with the original transcripts, as well as the context of each theme's codes and associated quotations. Refinement and consolidation of those initial themes occurred as a result of conversation between authors, which ultimately resulted in the final set of themes. Given the vast number of codes under each theme, some subordinate themes were developed that better accounted for nuanced ideas within the original transcripts. Those subordinate themes were synthesized in much the same way as the major themes, with every code in each major theme being grouped together and evaluated.

## 4. Findings

### *4.1. Descriptive Data*

Data analysis of the collected transcripts led to the creation of four themes and two sub-themes. Interviews included personnel from fire departments (n = 3), EMS agencies (n = 3), and emergency management departments (n = 2), with the eight agencies being distributed across the western (n = 5) and southern (n = 3) U.S. census regions. Law enforcement personnel were invited, but contacts who did respond displayed a hesitance to share operational details due to security concerns.

Each participant shared basic information about their agency and its jurisdiction, displayed in Table 5. The population of each agency's response area ranged from approximately 200,000 to 10 million people. The average call volume of the interviewees' agencies ranged from 36,000 to 200,000 calls per year. Reflecting the wide range of population and call volume among participants, there was substantial heterogeneity in each agency's operating environment. Some participants illustrated dense urban cities while others reported sparse metropolitan areas or interspersed natural landscapes. While some interviewees described being resource saturated—meaning additional first responders could be readily called upon from neighboring jurisdictions—others stated that their systems were relatively isolated. These differences and others suggest that the resulting themes, which were prevalent across the eight interviews and are discussed in detail below, are not restricted to a particular system, agency, or environment.

**Table 5. Interviewee descriptive statistics**

| Interview | Agency Type | Census Region | Population (Estimated) | Call Volume (if provided) |
|---|---|---|---|---|
| N1 | Emergency Management | West | 2,000,000 | |
| N2 | Fire/EMS | South | 650,000 | |
| N3 | Fire | West | 1,000,000 | 200,000 |
| N4 | EMS | West | 700,000 | 36,000 |
| N5 | EMS | South | 350,000 | 70,000 |
| N6 | Emergency Management | West | 10,000,000 | |
| N7 | EMS | South | 200,000 | 70,000 |
| N8 | Fire | West | 300,000 | |

### *4.2. Theme 1 – Continuity in the Face of Technical Disruptions*

Continuity of operations emerged as a significant focus of emergency preparation and contingency planning within ESS. Every participant indicated that their agency had some form of operational continuity planning and infrastructure in place, although the form varied between alternatives such as continuity of operation plans (COOPs), pen-and-paper backups, and Primary, Alternative, Contingency, Emergency (PACE) communication plans. One ensuing implication is that the continuity-oriented sector might prioritize availability over the other two aspects of the CIA triad. A few interviewees [N2, N5] explained the reason behind this phenomenon: the immediate mission of every emergency service is focused on *"life safety;"* even if they must pursue it in a reduced capacity, first responders are intent on ensuring that they maintain some way to respond to emergencies.

One participant [N5] expanded on that observation by comparing the presented scenario to their agency's experience during a major natural disaster. The incident had resulted in a complete breakdown of technical infrastructure and caused most communication devices to become non-functional. In the face of such a complete disturbance, the agency reportedly maintained its *"mission to rescue people,"* with the interviewee describing how their fellow responders *"just went out, found people, and rescued them. You know, there was no calling for help."*

Such analogies—comparing technical disruptions of a benign origin and those created with malicious intent—were common among interviews. Some participants [N1, N3, N7] reported experiencing 911 outages with their agencies in the past, leading them to feel prepared if the same type of incident were to reoccur. They explained that their response to a TDoS attack would consequently be similar, coping with a loss of 911 in the same way regardless of the actual cause. Similarly, several interviewees [N1, N5, N6, N7] pointed to their COOP/PACE plans and experience with temporary losses of radio communication to illustrate their belief in being able to effectively counter radio jamming. It became evident that, regardless of whether or not a technical disruption was malicious in nature, first responders would react to and manage it in practically the same way. The only difference, as one participant [N2] noted, is that a malicious TDoS attack would likely heighten their suspicion that *"something might be brewing,"* alluding to the role that situational awareness can play even before a complex incident begins.

Some interviewees [N1, N3] reported having historically experienced swatting incidents with their agencies. Those participants explained that they would not only feel personally comfortable confronting that aspect of the scenario, but that their agency would be well-equipped to handle it as well. Meanwhile, others [N5, N6] indicated that they could not recall historical instances of swatting with their agency, but that it nevertheless represented an attack-vector that would be easily resolved under current response procedures, aligning with the researcher designation of swatting as low-moderate severity. Swatting therefore largely matched researcher expectations of being both a common and readily addressable aspect of the scenario, validating its risk rating in Figure 1.

Across several interviews, a number of contingency plans and alternative operational capabilities were discussed, particularly for disruptions to internal agency communications.

Almost all participants (n = 7) reported being well-versed in handling instances in which MDTs are unavailable or dispatchers are unable to digitally assign responders to incidents. Some discussed an existing capacity for dispatchers to transition to *"pen-and-paper"* strategies (n = 6), while others described having temporary or permanent agency liaisons assigned to the local PSAP (n = 3). Multiple participants also cited the use of formal planning frameworks, including COOP plans (n = 4) for operations and PACE plans (n = 4) for communications. In either case, an emphasis was placed on the ability of the agency and its partners to rapidly transition following a technical disruption.

*4.3. Theme 2 – Planning for Failure*

Some participants [N2, N3, N6, N8] indicated that they and their agencies had engaged in planning for some worst-case scenario, although fewer (n = 2) had actually implemented formal procedures as a result of those conversations. Their prescribed reactions to such a scenario varied, with some relying on mutual aid and partner agencies (n = 2), others set on determining ad-hoc solutions should the situation arise (n = 2), and one entrusting frontline personnel to appropriately respond (n = 1). Notably, one participant [N3] described a hesitancy among their peers to engage in formal planning for a worst-case scenario:

> *"I've spoken to people where it's like, "hey, what if this happens?" "Well, we'll get through it. It'll be a little messy, but we'll get through it." And I'm like, "OK, well, where is that part written down?" Well, it's not written down and it's like, "OK, well, what if you're not here today?" . . . When it comes to long-term disasters or comes to the things that people don't want to talk about here, I think we have a long way to go."*

Another participant [N8] potentially offered an explanation for this hesitancy, stating that "when you just run the business day to day of a city running 35,000 calls and dealing with normal stuff that happens, it becomes difficult to go, *"oh, we're very busy and overwhelmed right now, but let's add this other giant project to our basket."*

Similarly, a few participants [N3, N4] indicated that despite having continuity plans in place, their leadership and frontline personnel do not regularly train on how to implement them. This contrasted with a number of interviewees [N3, N6, N7, N8] who reported regularly engaging in exercises related to terrorism and the single agency [N6] that had trained on managing technological disruptions during an incident—that interviewee described a mock scenario involving a cyber incident during a brush fire.

*4.4. Theme 3 – Frontline- vs. Command-oriented Approach to Technical Disruptions*

Diverging opinions existed among participants over the importance that extensive contingency planning plays in preparing frontline ESS personnel—the firefighters, police officers, EMS providers, and 911 dispatchers performing day-to-day emergency response—for a technical disruption or complex incident. A spectrum of perspectives emerged, spanning between two extremes that prioritized decision-making either at the command-level or at the frontline-personnel level.

A command-oriented approach to technical disruption emerged as being a preference to rely on procedures pre-written by command staff and those in emergency preparedness roles. This perspective was expressed by a few participants and was characterized by a focus on pre-planning, as well as a desire to ensure that all frontline personnel would know how to uniformly respond to a disruption or complex incident. One participant [N3] epitomized such an approach, explaining that a goal for their agency was to develop a procedural book that frontline and command personnel could *"pull off the wall,"* outlining a consistent reaction to 911 outages, loss of radio communications, or other technical disruptions. Another participant [N2] emphasized the importance of their COOP and PACE plans in the aftermath of a major technical disruption, though they simultaneously acknowledged the need to be able to adapt them to the unforeseen.

Contrasting with the command-oriented approach was a frontline-oriented one, denoted by a preference toward empowering frontline personnel to adapt to unprecedented emergencies and make the best decisions they can in pursuit of life safety. While discussing the scenario, one participant [N8] noted that their agency would be forced to flex and adapt in response to it because none of their procedures were prepared with the described scenario in mind: *"we don't have a lot of experience with that large of an incident. So we would be, I think, kind of forced to use what experience we have."* Another participant [N5] expressed a similar sentiment, explaining that the scenario would necessitate falling back on the experience that their personnel had derived from a tabletop exercise. Both respective agencies were said to maintain formal procedures, but that the agencies primarily used them as informal guidelines for frontline personnel's responses rather than as strict protocols to adhere to.

Underlying this distribution in strategy was a fear among some participants that frontline personnel lacked the requisite skills to effectively adapt to a sudden loss of technology. Multiple participants [N3, N4, N8] explained that newer frontline personnel were unfamiliar with the analog methods that were relied upon prior to the introduction of technology, therefore rendering them vulnerable to a loss of operational capability should they be forced to revert to the unfamiliar processes. Some of those participants further indicated that this problem was fueled by a loss of institutional knowledge, as veteran members were leaving the ESS workforce and taking memory of pre-digital processes with them. One interviewee highlighted this phenomenon by describing the immediate effect of transitioning to pen-and-paper in response to a CAD-based ransomware attack:

> *"A lot of these dispatchers, that's all they ever knew, is a computer aided system. When I started, we didn't have computers on the rigs. They went away from the old card system to a computer aided dispatch system. And so . . . I don't think there's a single dispatcher that was still working today that worked in the old system. So, it was somewhat foreign to them."* [N8]

Several participants [N3, N4, N8] shared similar anecdotes, illustrating the fact that agencies are increasingly composed of frontline personnel who lack analog skills. This represented a cornerstone for one participant's [N3] argument in favor of a command-oriented approach—*"the old school battalion chiefs, they were like 'whatever,' you know? And they just brought out the book with dust on it and were like, 'hey, this is how we did it in the old days, newbies.'"* While diverging opinions existed on the best approach for preparing for technical disruption, there remained agreement among participants that many frontline personnel were unprepared to quickly transition to analog.

### *4.5. Theme 4 – Agency Partnerships and Expertise*

ESS is known to be dependent on and interdependent with other critical infrastructure sectors. In addition, first responder agencies often respond alongside each other to accomplish separate mission functions during the same emergency. This was evident in the responses of most participants (n = 7), who discussed their agency's relationships with local emergency service partners, with other divisions of their governing bodies, and with third-party software providers. Opinions on the strength of these partnerships, as well as the benefits of maintaining them, varied among participants. Several discussed a lack of in-house technical expertise and operational difficulties they faced as a result of partnering with software providers, giving rise to subtheme 4.1. Partnerships with other first responders was also discussed, with law enforcement regularly being mentioned as a result of the research's focus on terrorism, leading to the development of subtheme 4.2.

#### *4.5.1. Subtheme 4.1 – Third-party outsourcing of technical expertise*

Several interviewees (n = 6) indicated that their agency outsources technical expertise to third-party government groups, namely municipal technology departments. Some interviewees viewed this as largely positive, describing how delegating technological management enables agency personnel to focus on emergency response and operational needs. One participant [N2] explained that a third-party municipal cybersecurity group regularly handled pushing system updates to computer systems, leaving agency IT specialists available to focus on needs specific to their agency: *"Our members really don't have to go in and do anything, and our IT work group here at the fire department doesn't have to push those updates or push any of that information out. It's all done in the background."* However, even with the benefits of being able to *"kick [cybersecurity alerts] over"* [N2], there were concerns that another participant emphasized while discussing their own agency's relationship with its municipal information services (IS) organization:

> *"We used to have our own IS people inside of our department that would actually work for us. [The municipality] wanted to consolidate it to their own, so they withdrew and now they just have them come in and do stuff for us. . . . That's why we have like John here now - he's not an IS or IT guy, but we really need somebody here that understands how the healthcare technology works and he's the guy that goes between us and IS and handles most of that stuff.*

*. . . Our day-to-day operations with [IS] is sometimes challenging because there's not enough of them here and they're not dedicated to us like they used to be."* [N7] [1]

Not only does outsourcing technology management risk a lack of specialized expertise for an agency, but it also removes constant and consistent access to generalized expertise. Participants were concerned about the impact that this separated organizational structure could have in cases of sudden critical technology failures.

This concern was visible in the reliance of agencies on third-party software, something that was corroborated by the transcripts: among the eight participants, 15 distinct cloud programs were identified as being used to coordinate staffing, inventory management, incident reporting, and other administrative tasks. While clearly beneficial for allowing first responders to focus on emergency response, failure of these technologies has historically created significant disruptions:

*"[Our staffing software provider] got hit about three years ago and they weren't gonna pay the ransom on it, which obviously affected the end users - which was us. But we did have some stuff in the back burner that we were actually getting ready to destroy in a couple months, just due to the retention policies. So, we were able to use that."* [N1]

In fact, half (n = 4) of the participants described their agency being negatively impacted by that same ransomware attack and similarly having to resort to pen-and-paper contingencies. Use of third-party software providers was described as beneficial, but was simultaneously noted to risk creating widespread disruptions that cannot be directly addressed by agency personnel.

*4.5.2. Subtheme 4.2 – Relationships with local partner agencies*

Every interviewee described the collaborative nature of their work and the close interagency relationships that they have cultivated, with many (n = 6) specifically detailing the implementation of automatic and mutual aid policies. Automatic aid refers to the immediate dispatch of additional resources from agencies with the same mission function in nearby jurisdictions [69]. One example might include two neighboring fire departments being simultaneously dispatched to a large fire based on a prior agreement. Mutual aid is a broader term that encompasses automatic aid and refers to any requests for assistance from neighboring agencies [69]. Positive standing relationships are critical to mutual aid, as one participant [N1] explained: *"We have great relationships with the counties that surround us. . . . So should something occur, we send resources to them."*

Strong relationships reportedly transcend the response phase for an agency: participants emphasized the importance of coordinated preparation for critical emergencies. Several interviews (n = 5) featured discussions of prewritten integrated response procedures, particularly in the case of active violence (AVIs) or mass-casualty incidents (MCIs). In discussing their response to both false reports and legitimate AVIs, participants described having created mirrored protocols [N3, N7] or executed joint trainings [N4, N5, N8] to ensure that frontline personnel were capable of a rapid, coordinated response. Two interviewees [N2, N5] also described intentional efforts to align their technology with partner agencies, with one [N2] explaining how "*the other agencies attend our quarterly [statewide interoperability coordinator] meetings and we talk about what is our PACE plan for interoperable communications*." Regular engagement and the facilitation of positive working relationships were evidently key to ensuring partner agencies could effectively work together in the field.

[1] Note that "John" is an alias used to maintain the study's commitment of anonymity for participants. Any similarity to real persons is purely coincidental.

As one participant noted while discussing the response of other agencies to a technical disruption, a highly collaborative approach to operations produces a positive externality:

> *"We have great working relationships and I think that's one of the key pieces that's missing in a lot of systems across the nation. They just don't have those kind of working relationships and it's very territorial. And so luckily, we all are pretty much on the same page on how we handle authentication and how we handle cybersecurity and redundancy platforms across the county."* [N4]

The "great working relationships" that had been created with partners for operational purposes led in turn to parallel relationships in the face of technical disruptions or cybersecurity failures. This phenomenon was not isolated either. Across multiple interviews [N1, N4, N6], it was stated that the same collaborative mindset dominating traditional emergency response had translated to the realm of cybersecurity and technical continuity of operations.

## 5. Discussion

The same interviewee responses that gave rise to the preceding themes also demand a return to the underlying questions motivating this research. They present previously unexplored implications for emergency preparedness and response among ESS, particularly in the face of an increasingly technical portfolio. Understanding continuity, planning approaches, and partnerships offers insight into preparation for cyber-based disruptions and the factors that could separate successful responses from their unsuccessful counterparts.

### *5.1. Response to a Simultaneous Technical Disruption and Terror Attack*

Theme 1 revealed that participating personnel felt largely confident in confronting individual cyber-based disruptions. Swatting was said by participants to be easily dealt with because first responders must commonly confront false alarms during daily operations. TDoS attacks and radio jamming, while potentially more impactful, were still comparable to historical 911 and communication outages, creating an experiential anchor point that participants felt comfortable falling back upon. However, diverging opinions exited as to whether it is preferred to allow frontline personnel to fall back on that experience and make decisions as needed, or whether command staff should be focused on developing reliable, accessible response procedures before a major disruption occurs. Of those participants who spoke on the subject, half (n = 4) landed in between the two extremes of frontline-oriented and command-oriented approaches, insisting that a balance was required between pre-planning and flexibility. The key determinant in either case is likely effective implementation. Adopting a command-oriented approach would require that ESS agencies facilitate recurring review of cyber-attack procedures and ensure their personnel are well-trained on such protocols. Meanwhile, a frontline-oriented approach would necessitate a workforce that is both technically literate and effective in analog operations. It follows that operating in between either extreme would require a combination of these strategies.

It is worth noting that this disagreement on the best strategy to adopt existed without the input of law enforcement, who did not participate in this study. This could have potentially shaped conclusions in a few ways, particularly in the scenario's context of a cyber-kinetic act of terror. Law enforcement often trains for AVIs and may therefore retain a comprehension of procedure and objectives in the absence of technological infrastructure, suggesting an ability to operate well in a command-oriented framework with preconceived strategies. Conversely, law enforcement officers also routinely encounter violence and scenarios that require improvisation to maintain public safety [70,71], suggesting they could also potentially operate well in a frontline-oriented framework. It is ultimately difficult to determine a non-speculative answer without input from law enforcement, a shortcoming that might be resolved with further work on this subject.

Another common assertion was the importance of partners and outside resources. Mutual aid and other first responder agencies, both inside and outside an agency's jurisdiction, were noted by participants to be vital to facilitating an effective response to complex kinetic attacks, as well as to suboptimal operations secondary to a technological disruption. This is largely unsurprising; first responders have access to different toolsets and accomplish different objectives, necessitating the ability to call on specialized resources and additional personnel. Those assets in turn can be used to supplement operational deficiencies during technology failures or localized disasters. That is why shared resources, interoperable communication systems, reconciled SOPs, and joint training exercises are valuable—they offer a mechanism for easily merging responses across shared geographies, multijurisdictional incidents, and mutual aid events.

It was made clear that cyber-based disruptions coupled with a kinetic attack would stress even the most advanced and progressive emergency response systems among respondents. Nevertheless, participants insisted that they would continue doing what they must, formulating ad-hoc responses to unprecedented challenges in order to effectuate a response. This suggests that preparation for technological disruptions would not be a question of whether ESS is capable of responding to or recovering from such an incident but rather represents a measure of the impact that the incident would have before resolution.

*5.2. Preparation for Technical Disruption*

Several noteworthy observations were made across the themes related to the preparation of respondents for complex incidents or technological disruptions. Themes 1 and 2 suggested that there were often formal procedures in place to ensure continuity of operations, such as COOP or PACE plans, yet there remained a tendency among participating agencies to avoid discussion of uncommon, worst-case scenarios that could occur. Rather than representing yet another failure of imagination, this phenomenon might better be referred to as a failure of priority. With a limited set of personnel to call upon and a strict necessity to ensure that day-to-day operations and emergency responses are handled, emergency planning was often deemphasized among participants. This was particularly glaring among those systems that lacked dedicated emergency management personnel. This lack of expertise and deprioritization of planning was justified by participants who pointed to the lower likelihood of a complex incident occurring—interviewees expressed uncertainty regarding whether incidents like the described scenario were worth planning for, rather than simply generating ad-hoc responses should they occur. While acknowledging the fiscal challenges that agencies continue to face [72], great benefit could be afforded by embedding dedicated emergency management personnel within non-EM agencies whose expertise could be invested into adequate disaster preparation.

Themes 3 and 4.1 also pointed toward a loss of technical knowledge within the interviewed segment of the ESS workforce, with the former suggesting a loss of institutional knowledge to retirement and the latter indicating a loss of specialized technical expertise. It was explained that, as veteran first responders reach retirement age or move out of operational positions, the ESS sector loses knowledge of how to conduct analog operations. Additionally, such veterans are replaced by members from younger generations who are more accustomed to performing work with technological assistance, which risks further hampering efforts to perform operations without such technology. Mark43 stated in their 2024 U.S. Public Safety Trends Report that "administrative duties such as paperwork and complying with local, state, and federal reporting requirements consume an increasing part of a first responder's day," which has led to the increased use of cloud programs for administrative purposes [73]. An increased use of specific technology and software combined with a loss of specialized expertise could risk leaving ESS even more vulnerable should their systems fail at a critical juncture. Considering that many agencies reported outsourcing technical and cybersecurity support to third-party companies and municipal departments, it is possible for a technological 'brain drain' to be occurring within the emergency services. Studying such a phenomenon's extent and impacts represents an interesting avenue for future study.

The preparedness of participating ESS agencies is best considered against the cyber maturity of other critical infrastructure. Several sectors including energy and chemical are governed by nationally standardized regulatory models, unlike ESS [74,75]. While those other sectors have grown to a high level of cyber maturity, participants noted that their agencies prefer ad-hoc approaches to cyber disruptions or lack appropriate personnel to manage an incident. This suggests the possibility of a reduced cyber maturity within the public safety sector.

It is impossible to assert based solely on this article's findings that ESS is fully prepared or unprepared for a sophisticated cyber-based incident. While the interviewed agencies were generally prepared for complex incidents and occasional technical failures, they appeared less prepared for enduring failures or a hybrid event. Participants were seemingly defined by an overreliance on technology that rendered them vulnerable in the case of any significant disruption. However, those same agencies generally remained unable to reverse the tide on overreliance because they lacked personnel who could focus on securing infrastructure or planning for worst-case scenarios while day-to-day operations are simultaneously maintained. In short, one of the current preparation challenges for ESS seems to be ensuring appropriate technology, cybersecurity, and emergency management knowledge across the roles of both experts and frontline personnel.

### *5.3. Factors Contributing to Differing Levels of Technical Disruption Preparedness*

A restricted sample size limits the conclusions that can be drawn about what distinguishes preparation between individual agencies. Nevertheless, some observations can be made based on the emerging themes. Theme 2 examines the fact that only some agencies engage in planning for 'worst-case' scenarios, which has led to planning failures based on failures of the imagination. This suggests a difference in mentality in which some agencies are more disposed to discuss unlikely incidents and are consequently more prepared should they occur, even if formal procedures are never implemented.

Themes 1 and 4.1 also indicated that preparation is in part built upon experience, with multiple participants describing feeling more prepared for aspects of the scenario based on similar historical incidents. As a result, different levels of preparation may also be based on different levels of exposure. Anecdotal evidence suggested that those who have responded to terror attacks before are likely to feel more comfortable doing so again in the future, and that the same stands for encountering cybersecurity incidents. Therefore, preparation may ultimately differ between agencies in part due to differences in planning mentality and experience.

## 6. Conclusion

Emergency personnel, researchers, and government agencies have long acknowledged the dangers associated with disruption of ESS technology and the negative impact that it can have on first responder operations. However, there has been little work on understanding the scale and scope of technological dependence among ESS agencies, and less focus still on quantifying the vulnerability it introduces. This research began addressing that gap and identified that, while prepared for complex kinetic incidents and short-term technological disruptions, participating agencies and personnel were not fully prepared for complex, long-term technical disruptions or hybrid cyber-kinetic attacks. This was contributed to by a loss of institutional knowledge to workforce retirement, overreliance by new personnel on technology, and the outsourcing of technical expertise to third parties. The findings suggest that first responders should continue to prioritize continuity efforts, ensure analog capabilities are not lost to technical innovation, and prioritize establishing technical and emergency planning expertise within their agency and across ESS as a whole.

These findings should be considered in light of the study's limited sample, which was concentrated among metropolitan agencies in the Southern and Western United States and did not comprehensively represent all segments of ESS. The study's intentional focus on agencies more likely to experience cyber- or kinetic attacks may also limit the applicability of these findings to austere, non-professional, or under-resourced agencies. Given the preliminary nature of this work, opportunity remains for further research to address questions raised by these conclusions. Research into the benefits of frontline versus command-oriented approaches could help understand the most effective process for continuity planning among ESS. An assessment of disparities in emergency and cybersecurity preparedness between metropolitan and austere ESS agencies could help generalize this research's findings beyond busy urban systems. Finally, a quantification of the entire sector's cybersecurity posture could identify common weaknesses and attack surfaces. With more than 78,000 emergency service agencies across the United States [76], there remains much to learn in understanding the vulnerability of ESS to cyber-attacks as the field continues to technologize.

**Funding Sources**

This research did not receive any specific grant from funding agencies in the public, commercial, or not-for-profit sectors.

**CRediT authorship statement**

**James Connor:** Conceptualization, Data curation, Formal analysis, Investigation, Methodology, Project administration, Visualization, Writing - Original Draft, Writing - Review & Editing. **Courtney Crooks:** Methodology, Validation, Writing - Review & Editing. **Ryan Shandler:** Conceptualization, Project administration, Supervision, Writing - Review & Editing.

**Acknowledgements**

The authors are grateful to Dr. Sy Goodman for his helpful discussions in the early stages of ideating this study. The authors also thank Trevin Hunter for his sharing of information related to Operation Red Dawn.

**Appendix A. Justifications for risk ratings of ESS attack vectors**

Table A.1 displays the criteria used to determine risk subcategorizations across both likelihood and severity. These criteria were then used alongside the justifications in Table A.2 to create the final risk matrix depicted in the article (Figure 1).

**Table A.1: Likelihood and severity scales for cyber-attacks targeting ESS**

| Rating | Likelihood Description | Severity Description |
|---|---|---|
| Low | Possible in a controlled environment but only seen a handful of times in the wild, if at all | Little to no disruption on the ability of emergency services to respond or conduct operations |
| Low-Moderate | Occasionally observed in the wild | A potential for disruption exists that is inherently limited by some factor, such as time or geography. |
| Moderate | Observed multiple times in the wild, but requires high technical expertise | Widespread disruption to normal operations would occur, but emergency services would retain their ability to respond to emergencies. |
| Moderate-High | Observed multiple times and requires limited expertise | Widespread disruption would occur, and could only be countered by the use of external resources. |
| High | Observed multiple times and requires little to no expertise | Widespread disruption would occur with the potential for a complete breakdown in operational capability. |

**Table A.2: Justifications of likelihood and severity ratings by attack vector**

| Attack Vector | Likelihood Justification | Severity Justification |
|---|---|---|
| Radio Eavesdropping | <u>Low</u>: The TETRA and P25 radio encryption standards are known to be vulnerable to compromise [25,77]. While eavesdropping on insecure communications has been exploited by criminal actors [78], the same cannot be said for encrypted systems. | <u>Moderate</u>: Eavesdropping by criminal actors would not undermine the ability to respond to emergencies. However, it could potentially limit what information can be shared via radio or enable targeting/avoidance of first responders. |
| Radio Forgery | <u>Low</u>: See radio eavesdropping. | <u>High</u>: Forging radio communications could enable a malicious actor to dispatch responders to false emergencies, inhibit situational awareness, and disrupt the ability of emergency responders to operate with a ground truth for incidents. It would significantly undermine the ability to appropriate identify, respond to, and manage incidents. |
| Ransomware on CAD | Low-Moderate: Ransomware attacks against CAD are known to occasionally occur, with 9 incidents in 2023 and 18 in 2024 [21]. | Moderate: While interruptions to normal operations would occur, workarounds are typically in place to manage CAD failures [21]. |
| Radio Jamming | Low: Radio jamming has been occasionally observed by real first responders, but has yet to become a quantifiably common practice and requires technical expertise [79–81]. | Moderate-High: Radio frequency interference could cause a loss of situational awareness among first responders and possibly interfere with their ability to be dispatched to emergencies. It could be countered through the use of alternative frequencies, networks, or communication devices. |
| Data Breach | Moderate: Data breaches regularly occur for first responder agencies. Many data breaches occur because of human error [82]; however, breaches being considered under this research are predicated on malicious, purposeful activity, which requires more sophisticated technical expertise to successfully execute. | Low-Moderate: Data breaches do not present an explicit disruption to emergency responses and are largely unimpactful on the ability of first responders to conduct operations. They may cause emotional harms among ESS personnel and thereby inhibit their response capacity, but such harm is limited by the type of information that is released. |
| Ransomware on PSAP | Moderate: Ransomware attacks against PSAPs are known to occasionally occur, with two incidents in 2023 and four incidents in 2024 [20]. | Moderate: Emergency service agencies would retain their ability to respond. With backup systems or secondary PSAPs in place, the interruption to critical communications infrastructure might be minimized. However, there remains the potential for communications to be disrupted between responders and with 911 callers. |
| Large TDoS Attack | Moderate: TDoS attacks have historically occurred, but require access to a significant amount of computing and telephonic resources to successfully execute (e.g. a botnet) [19]. | Moderate-High: DHS has described the extreme degradation effects that a TDoS attack can have on emergency communications, which can often only be mitigated through the use of external resources [83]. |
| Website Defacement | Moderate: While it has been demonstrated historically, defacement of a targeted website often requires moderate technical expertise in order to identify an appropriate vulnerability and successfully exploit it. | Low: Defacement of an emergency service agency's website would have no direct impact on their ability to appropriately respond to emergencies. |
| Data Breach of LEO PII | Moderate-High: Like typical data breaches, a certain elevated level of technical expertise is required to carry out a breach targeted toward law enforcement. However, law enforcement tends to be a common target for data breaches as a result of motivated targeting by hacktivists and cybercriminals [84]. | Moderate: Whereas a typical data breach tends to have a lesser impact on operations, widespread revelation of the personal information of law enforcement could lead to officers calling out of work, fearing for the safety of themselves or their family, or being targeted by swatting. |
| EAS Compromise | Moderate: Multiple instances of EAS compromise have been documented, including the false activation of siren-, television-, and SMS-based systems [85–87]. | Low-Moderate: while disruptions might occur for 911 call-takers, ultimately the primary effect of EAS compromise is to induce public panic and potentially disrupt the ability of first responders to distribute credible information during a major incident. As a result, operational disruption is inherently limited to major incidents. |
| Swatting | High: Swatting has been seen commonly across jurisdictions, with incidents seemingly increasing in frequency over the past decade. Execution of a swatting attack requires little to no technical expertise, except for the ability to spoof a caller ID, which can be done using online services [88]. | Low-Moderate: While there might be a temporary disruption to operations while a significant number of resources responds to the incident, that disruption is limited to the amount of time it takes for the initial responders to determine the unfounded nature of the report. |

## Appendix B. Full Interview Scenario

*General Description*

This scenario outlines a hypothetical terror attack by an unspecified threat actor (TA) group. In the scenario, agents of the TA group launch a series of coordinated physical attacks. Those attacks are preempted by cyber-attacks targeting first responders and emergency service infrastructure. Including false 911 calls, denial-of-service attacks, and the jamming of radio communications in the vicinity of the physical attacks.

*Detailed Scenario*

The attack begins with the placement of multiple false calls to 911 by agents of the TA group. Those calls report active shooters at two distinct synagogues in the region. The calls are made with the intent of distracting law enforcement and other first responders, while simultaneously displacing them away from the intended target.

Minutes after the false 911 calls are placed, a telephony denial-of-service (TDoS) attack is launched against the city's public safety answering point (PSAP). Dispatchers within the PSAP receive calls that, when answered, immediately hangup or simply remain silent until the dispatcher terminates the call. The calls arrive at such a high rate that they crowd out legitimate callers, who are placed into a queue or are forced to redial 911. The false calls additionally propagate over to secondary PSAPs, overwhelming the existing contingencies that are used to address unusually high call volumes.

Minutes after the sustained TDoS attack begins, the TA group launches the physical attack. Three targets are selected and attacked in a synchronized manner.

In the weeks prior to the physical attack, agents of the TA group had placed radio jamming equipment in an apartment that was relatively equidistant to each of the three physical attack locations. While sparse testing may have been conducted, the jamming equipment was only activated in a sustained manner at the time of the physical attacks. The equipment disrupts radio communications for all local emergency service agencies in the immediate area.

Emergency calls from victims of the physical attacks are delayed in reaching the PSAP due to the ongoing TDoS attack. Once reports are received, a significant response from first responders is delayed due to the displacement of emergency services from the false 911 calls. Upon arrival in the areas, first responders and emergency management personnel face challenges establishing incident command or developing a coordinated response due to radio jamming. The delays caused by these cyber-support mechanisms will later be determined to have created an increased casualty count from the physical attacks.

**Appendix C. Interview Questionnaire**

- Organizational questions
  - What is your title and what are the primary responsibilities of that position?
  - Describe your agency and what it does.
  - Approximately how large is your response area/jurisdiction?
  - Approximately how many people are in your response area/jurisdiction?
  - Approximately how many personnel does your team/agency employ?
  - Which of the following best apply to you? [Fire-Rescue w/o EMS, Police Department, EMS, Fire-Rescue w/ EMS, Emergency Management, Public Works]
- Use of/Reliance on Technology by the Emergency Services
  - What forms of technology does your agency use?
  - How prepared is your agency for a disruption of technology?
  - How prepared is the wider emergency response system prepared for a disruption in technology?
  - Can you describe any instances in which your agency has had to adapt due to a loss of technological capabilities?
- Preparation for Physical Attacks
  - What experience does your agency have in responding to physical attacks?
- Preparation for Cyber-Attacks
  - How has your agency approached cybersecurity in the past?
  - If applicable, can you describe previous cyber-attacks/incidents that your agency has dealt with?
  - What attempts at mitigation has your agency adopted since those occurred?
- The Attack Scenario (asked after reading the scenario in full)
  - How well do you feel your agency would be able to respond in such a scenario?
  - Has your agency dealt with instances of "swatting" or false 911 calls before? How have those calls impacted your agency?
  - How would your agency respond to a situation in which 911 calls are being disrupted?
  - How would your agency be impacted by radio jamming equipment being used?